\documentclass{elsart}
\usepackage{graphicx}

\usepackage{amssymb}

\begin{document}

\begin{frontmatter}

\title{Collective Current Rectification}

\author{S. Denisov}
\ead{sergey.denisov@physik.uni-augsburg.de}

\address{Institute of Physics, University of Augsburg,
Universitatsstrasse 1, D-86159 Augsburg, Germany}

\begin{abstract}
We consider a set of coupled underdamped ac-driven dynamical units
exposed to a heat bath. The coupling scheme defines the
absence/presence of certain symmetries, which in turn cause a
nonzero/zero value of a mean dc-output. We discuss dynamical
mechanisms of a dc-current appearance and identify current reversals
with synchronization/desynchronization transitions in the collective
ratchet's dynamics.
\end{abstract}

\end{frontmatter}

\section{Introduction}
\label{1}

The \textit{ratchet} effect, i. e. the possibility to obtain the
directed transport by using zero-mean perturbations only, has
induced notable scrutiny over the last decade [1]. Initially, most
studies have been focused on noisy overdamped models  have been
inspired by a molecular motors realm [1]. Then the ratchet's
approach has been applied to a broad class of physical systems in
which inertia effects are essential [2]. The examples are Josepshon
junctions [3], cold atoms systems [4], and mechanical engines [5].

Recently it has been shown that the ratchet idea can be viewed as a
part of a general symmetry-breaking approach [6]. This approach is
based on the analysis of all relevant symmetries which have to be
broken in order to fulfill necessary conditions for a dc-output
appearance. The formalization of symmetry analysis for one
particle's dynamics has been addressed in Ref.[7].

In the present paper we aim at  collective rectification effects
which arise in a set of coupled single ratchet units. Although
various examples of interacting ratchet systems have been proposed
already in the context of molecular motors [8], here the emphasis is
put on the weak-noise underdamped limit. We show that a coupling
scheme determines a set of certain symmetries for the ratchet's
collective and hence defines necessary conditions for a dc-current
generation. Dynamical mechanisms of a current rectification are
connected with a coherence between units, which depends not only on
a coupling scheme, but also on a strength of interactions.

\section{Coupling schemes and symmetries}
\label{} Let us consider a set consisting of $N$ identical dynamical
units, $\mathbf{x}=\{x_{i}, i=1,...,N\}$, that are linearly and
symmetrically coupled. The coupling scheme is described by some
graph which can be encoded in the symmetrical $N\times N$ binary
matrix $G_{ij}$, $G_{ij}=G_{ji}$. The equations of motion are the
following:
\begin{eqnarray}
\ddot{x}_{i}=-\alpha\dot{x}_{i}+F(x_{i}-x_{i}^{0},t-t_{i}^{0})+
c\sum_{j=1}^{N}G_{ij}H(x_{i},x_{j},\dot{x}_{i},\dot{x}_{j})+\xi_{i}(t),
\end{eqnarray}
where $F(x+L,t)=F(x,t+T)=F(x,t)$ is the double periodic force
function, $H$ is linear over all arguments coupling function, and
$c$ is the strength of interactions. The stochastic terms
$\xi_{i}(t)$ are mutually independent delta-correlated Gaussian
white noises, $\langle \xi_{i}(t)\xi_{j}(s) \rangle=\sigma
\delta_{ij}\delta(t-s)$, where $\sigma$ is the noise intensity. We
assume also that  the force function, $F$, is the same for all the
units, but the phases, $x_{i}^{0}$ and $t_{i}^{0}$, can be different
for different units. Finally, we are interested in the mean
dc-current,
\begin{equation}\label{2}
J=\lim_{t\rightarrow \infty} \frac{1}{Nt}\sum_{i=1}^{N}x_{i}(t).
\end{equation}

Following  the symmetry analysis ideology [6-7], in order to
determine  necessary conditions for a dc-output appearance, we have
to check whether there exist symmetry transformations  which allow
to generate out for each trajectory of the system (1) another one
with a reversal velocity. The presence of the white noise,
$\xi_{i}(t)$, does not change the symmetry properties of the system.
Moreover, the coupling to a heat bath leads to an effective
exploration of the whole phase space and produce an averaging in a
case of several coexisting attractors [6,9].

For the one-particle case, $N=1$, a transformation of interest has
to involve a change of the sign of $x$ (and thus change of the
current direction). It allows also some shifts in time and space
domains [6-7]:
\begin{equation}\label{3}
\hat{S}_{single}: x\rightarrow-x+\lambda,~~~t\rightarrow t
+\tau,~~~~x\in \textbf{R} .
\end{equation}

In the case of several coupled ratchets a symmetry operation should
be performed in the global coordinate space $\textbf{R}^{N}$ and can
also involve a permutation between different units. The
corresponding symmetry operation can be described as the linear
transformation,
\begin{equation}\label{2}
\hat{S}_{network}:
\mathbf{x}\rightarrow-\mathbf{S}\mathbf{x}+\mathbf{\lambda},~~~
t\rightarrow t+\tau,~~~~\mathbf{x}\in \textbf{R}^{N},
\end{equation}
where $S_{ij}$ is a $N\times N$ binary matrix with only one non-zero
element at each row and line, and $\mathbf{\lambda}=\{\lambda_{i}\},
i=1,...,N$ is the vector of shifts. The matrix $\mathbf{S}$ encodes
permutation between units.

The properties of the permutation matrix $S_{ij}$ (and the existence
of such a matrix at all) are strongly depend on the structure of
connections among  units. Below, using simple examples, we will
illustrate this statement.

{\it Breaking symmetries by connections.} Let us first consider the
case of  two particles (rotators) in the standing-wave potential
with the modulated amplitude [10], coupled by the linear spring:
\begin{eqnarray} \label{1}
\ddot{x}_{1}=-\alpha\dot{x}_{1}+F(x_{1},t)+c(x_{2}-x_{1})+\xi_{1}(t)
\\
\ddot{x}_{2}=-\alpha\dot{x}_{2}+F(x_{2}-x_{0},t-t_{0})+c(x_{1}-x_{2})+\xi_{2}(t),
\end{eqnarray}
where $F(x,t)=\sin(x)\sin(\omega t)$.

In the uncoupled case, $c=0$, both the systems posses symmetries of
the type (3):
\begin{eqnarray}\label{2}
\hat{S}_{1}: x\rightarrow-x+ \pi, t\rightarrow t+T/2 \\
\hat{S}_{2}: x\rightarrow-x+\pi+x_{0}, t\rightarrow t+T/2+t_{0},
\end{eqnarray}
with $x_{0}=t_{0}=0$ for the first rotator, Eq.(5). The symmetry
transformations, Eqs.(7-8), are independent for each rotators. The
whole system, Eqs.(5-6), is symmetric with respect to the
transformation $\hat{S}_{1} \times \hat{S}_{2}$ for any choice of
$t_{0}$ and $x_{0}$. So, the mean dc-output for the uncoupled case
is zero (line(1) in Fig.1a).

In the  case of coupled particles, $c>0$, abovementioned
transformations, , Eqs.(7-8), should been conjugated and independent
spatial and temporal shifts are forbidden now. For $x_{0}\neq k\pi$
and $t_{0}\neq T/2$, the connection breaks both the symmetries and
we can expect on nonzero current appearance (see also Ref. [11] for
another example of an overdamped dimer with an additive driving
force). For the set of parameters $\alpha=0.1$, $w=0.3$,
$x_{0}=\pi/2$, $\sigma=0.01$ and $t_{0}=T/4$, the symmetry violation
is realized by the asymmetrical limit cycle with the negative
winding number (line(2) in Fig.1a). So, in this case  the connection
between the units destroy all the symmetries and leads to the
dc-current generation.

\begin{figure}[htbp]
\centering{\resizebox{8cm}{!}{\includegraphics{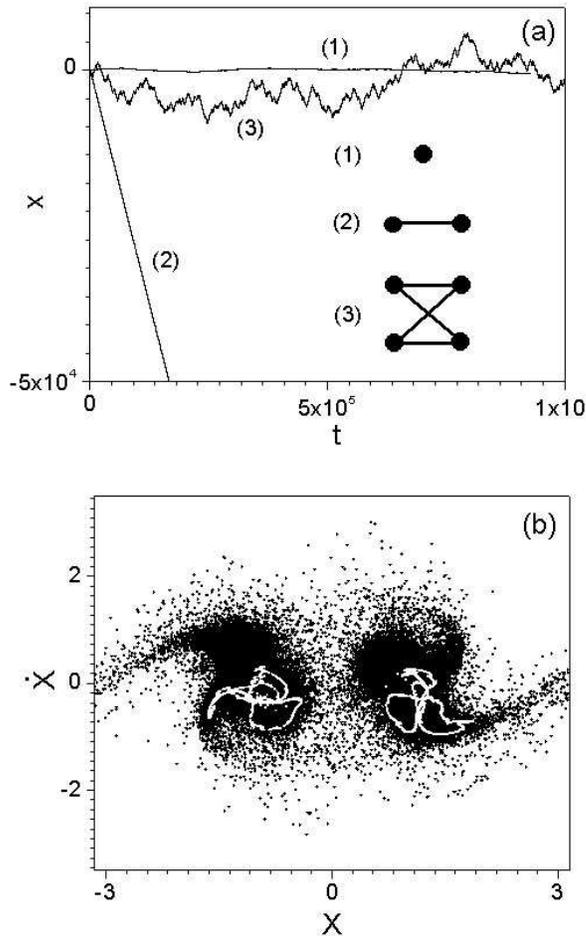}}}
\caption{(a) The dependence of $x_{1}(t)$ versus $t$ for different
coupling schemes (see text for details) for the parameter values
$\alpha=0.1$, $w=0.3$, $x_{0}=\pi/2$ and $t_{0}=T/4$ and noise
intensity $\sigma=0.01$. The dependencies for another rotators from
the set are the same due to the homogeneity of the system, Eqs.(5-6)
and Eqs.(9-10); (b) Poincar\`{e} sections for the first rotator for
the third variant of the coupling scheme. The coupling to the heat
bath leads to the averaging over two symmetry-related chaotic
attractors (white dots) with opposite mean velocities.}
\end{figure}

{\it Restoring symmetries by connections.} Let us now consider the
dimer  identical to the previous one, Eq.(5-6), but spatially
shifted by $\pi$ (half of the period):
\begin{eqnarray} \label{1}
\ddot{x}_{3}=-\alpha\dot{x}_{3}+F(x_{3}+\pi,
t)+c(x_{4}-x_{3})+\xi_{3}(t)
\\ \ddot{x}_{4}=-\alpha\dot{x}_{4}+F(x_{2}-x_{0}+\pi, t-t_{0})+c(x_{3}-x_{4})+\xi_{4}(t).
\end{eqnarray}
Both the systems, Eqs.(5-6) and Eqs.(9-10), can be transformed from
one to another by the simple coordinate shift. Thus, the system in
Eqs.(9-10) produces the same mean current, as the system in
Eqs.(6-7).

As the next step we couple both the systems by  additional
connections $((1)\leftrightarrow (4), (2) \leftrightarrow (3))$ (see
the inset in Fig.1a). The generalized symmetry transformation,
Eq.(4), can be identified with the following permutation matrix:
\begin{equation}
S_{ij}=\left(\begin{array}{cccc}1&0&0&0\\0&0&0&1\\0&0&1&0\\0&1&0&0
\end{array}\right),
\end{equation}
and the shift vector $\mathbf{\lambda}=\{\pi\}$.

Here,  despite  to the previous case, the introduction of additional
connections leads to the appearance of the new symmetry and we may
expect on the nullification of the dc-output (line(3) in Fig.1a).
For the above set of parameters we found that the symmetry is
realized in the phase space by  two symmetry-coupled chaotic
attractors with opposite mean velocities (see Fig.1b). It is easy to
check, that any other type of connection with two additional links
does not restore the system's symmetry.

\section{Synchronization and current reversals}
\label{}

While the presence or the absence of the dc-output is clearly
connected to the absence/presence of the  symmetry transformation,
the dc-current value is determined by dynamical mechanisms. From the
previous studies [12] it is known that global  properties of a
collective dynamics are defined by a coherence between units. Thus,
we can expect that an efficiency of the current rectification is
closely related to a degree of synchronization within a ratchet's
collective.

As a natural example we consider the model of $N$ globally coupled
array of underdamped Josepshon junctions (JJ), subjected to an
ac-current $E(t)$ [13]. The equation for the superconducting phase
difference $x_{i}$ across the single junction is
\begin{eqnarray} \label{1}
\ddot{x}_{i}=-\alpha\dot{x}_{i}+sin(x_{i})+E(t)+
c(\dot{x}_{i}-\langle \dot{x} \rangle_{A})+\xi_{i}(t) ,
\end{eqnarray}
where $\langle \dot{x} \rangle_{A}=\frac{1}{N}
\sum_{i=1}^{N}\dot{x}_{i}(t)$ is the instantaneous mean array
current. As the driving force we used the two-harmonics combination,
$E(t)=cos(\omega t)+cos(\omega t+\pi/2)$, which ensures that all the
relevant symmetries are broken [7]. The numerically obtained
dependence of the array dc-output on the strength of interaction $c$
is shown in Fig.2(a). For the set of parameters $\alpha=0.1,
\omega=1,  \sigma=0$, the dependence demonstrates the presence of
two successive current reversals, at $c \approx 0.03$ and $c\approx
0.33$. In order to understand dynamical mechanisms of these events,
we introduced the mean decoherence, which we define as
\begin{equation}
D=\langle | \dot{x(t)}-\langle \dot{x} \rangle_{A}| \rangle_{A,T},
\end{equation}
where $\langle ... \rangle_{A, T}$ means  the averaging over the
array and over the one period of ac-driving. The first result from
the comparison of both the dependencies (Fig.2(b))is that the
current reversal at $c \approx 0.33$ is connected with the
transition to  the regime of the complete synchronization, $D=0$.

\begin{figure}[t]
\centering{\resizebox{8cm}{!}{\includegraphics{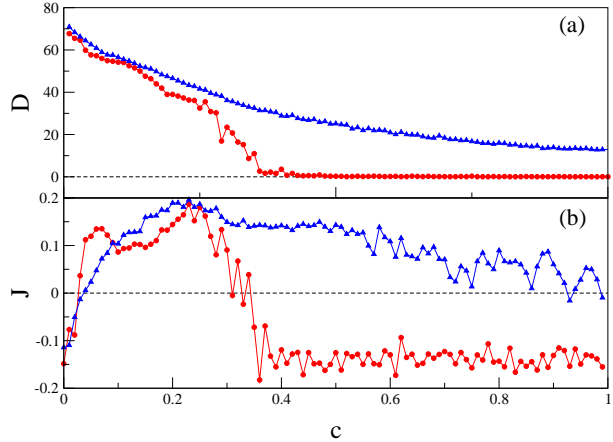}}}
\caption{The dependence of the mean dc-output, $J$, and the degree
of decoherence, $D$,  versus the strength of interaction $c$ for the
system of $N=50$ coupled units  from Eq.(12). The values of
parameters are $\alpha=0.1$, $w=1$, $\sigma=0$ (circles) and
$\sigma=0.03$ (triangles).}
\end{figure}

Lets now track briefly a relation between the current reversal and
the synchronization. Initially, in the uncoupled limit, $c=0$, all
JJ's work independently, each one as the single one-dimensional
rectificator [1,6]. The asymmetry of the ac-force $E(t)$ is realized
through the chaotic attractor with the positive mean current (see
Fig.3a) [14]. The local phase space of each units, $(x_{i},
\dot{x}_{i},t)$, has identical structure, but the relative phases of
different units are randomly distributed (depending on the initial
conditions).

After the introduction of a nonzero coupling, $c>0$, some coherence
between the units  occurs.  From a point of view of the dynamics of
the single junction, this leads to changes of an attractor structure
in the phase space $\textbf{R}^{3}$ (see Fig.3b). This causes
changes in a projection of the attractor's invariant density on the
velocity subspace $\dot{x}_{i}$ and, as a result, leads to the
current reversal at $c\approx 0.03$. The further increase of the
interaction strength up to $c \approx 0.33$ results in the complete
synchronization and to a shrinking of the global attractor in
$\textbf{R}^{2N+1}$ to the hyperplane
$\textbf{R}^{3}$,$(x_{i}=x,\dot{x}_{i}=\dot{x},t)$. The global
attractor has now dimension equal to $3$. The attractor of the
single unit now is the same as in the uncoupled limit but all the
units have the same phase. The mean dc-output returns to its value
at the limit $c=0$.

\begin{figure}[t]
\centering{\resizebox{10cm}{!}{\includegraphics{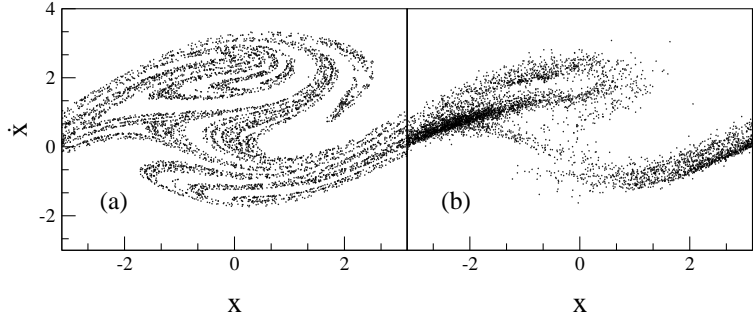}}}
\caption{Poincar$\grave{e}$ sections for a single junction from
the array of $N=50$ coupled JJ, Eq.(12), for (a) $c=0$ and (b)
$c=0.27$ (zero-noise case).  Parameter values are the same as in
Fig.2.}
\end{figure}

Here some analogy with a current reversal in an one-dimensional
deterministic ratchet [14] can be drawn. In the one-dimensional case
current reversals have been identified with tangent bifurcations
from chaotic to regular (limit cycles) attractors [14]. In  our case
the current reversal at $c \approx 0.33$ corresponds to the
transition "hyperchaos - chaos" [15], connected with the shrinking
of a the system attractor in a global phase space
$\textbf{R}^{2N+1}$. It is interesting to note, that from a point of
view of the local attractor of the single junction, this transition
corresponds to a crisis, connected with {\it expanding} of the
attractor in three-dimensional subspace $(x_{i}, \dot{x}_{i},t)$
(compare Fig.3a and Fig.3b).

The presence of a weak noise (which is equivalent to a weak coupling
with a heat bath) can strongly suppress correlations between units
and may lead to a delayed synchronization transition [16]. Due to a
strong conjugation between the synchronization and the current
rectification process, we can manage a noise-induced current
reversal for a  fixed strength of interaction (Fig.2(b)). Thus, the
presence of a thermostat allows to control the dc-output by changing
a temperature of the system.

\section{Concluding remarks}
\label{}
 Finally, we have presented the symmetry approach to the problem of
the collective current rectification by a set of coupled dynamical
units. This idea can be used for a more general problem such as an
obtaining of a non-zero value of some relevant mean ensemble
characteristic, $\langle A(\mathbf{x})\rangle_{t}$. This
characteristic can correspond, for example, to a mean magnetization
of a spin lattice with a complex geometry [17]. The proposed
collective ratchet's ideology may also be relevant in a context of a
cooperative dynamics of neural networks, as an approach to a visual
processing with directional selectivity [18].

Relations between  symmetries of  a collective ratchet and its
coupling scheme on the one side, and relations between a topology of
interactions and synchronization properties [19], on the other one,
may open an interesting perspective. It has been found that in the
thermodynamic limit, $N \rightarrow \infty$, the synchronization is
impossible for  nearest-neighbor coupled dynamical networks if the
number of sites connected to a given site, $N_{c}$, is a finite
fraction of the total number of sites $N$, $\frac{N_{c}}{N}=const$
[20]. On the other hand, in a case of small-world networks, i.e.
sets with  long (in a sense of a topological distance) connections,
the synchronization can be achieved in the thermodynamic limit
through a small fraction of distant connections [20]. This
nontrivial effect provides a tool for a control of the dc-output in
a massive collective ratchet by changing a small number of relevant
connections only.

This work has been supported by the Emmy Noether-Programm of the DFG
under contract LU1382/1-1

\end{document}